\documentclass[]{osa-custom}
\usepackage[utf8]{inputenc}
\usepackage[T1]{fontenc}
\usepackage{graphicx}
\usepackage{grffile}
\usepackage{longtable}
\usepackage{wrapfig}
\usepackage{rotating}
\usepackage[normalem]{ulem}
\usepackage{amsmath}
\usepackage{textcomp}
\usepackage{amssymb}
\usepackage{capt-of}
\usepackage{hyperref}
\journal{oe}
\articletype{Research Article}
\usepackage{epstopdf}
\usepackage{subcaption}
\usepackage{physics}
\usepackage{float}
\usepackage{textcomp}
\let\origeqref\eqref
\newcommand{\eqrefn}[1]{Eq. \origeqref{#1}}

\newcommand{\lcwl}{\lambda_{\text{cwl}}}

\newcommand{\cra}{\theta_{\text{CRA}}}

\usepackage{tikz}
\date{\today}
\hypersetup{
 pdfauthor={Thomas Goossens},
 pdftitle={},
 pdfkeywords={},
 pdfsubject={},
 pdfcreator={Emacs 26.3 (Org mode 9.2.6)}, 
 pdflang={English}}
\begin{document}

\title{A vignetting advantage for thin-film filter arrays in hyperspectral cameras}
\author{Thomas Goossens\authormark{1,2,*}, Bert Geelen\authormark{2}, Andy Lambrechts\authormark{2}, Chris Van Hoof\authormark{2,1}}
\address{\authormark{1}Department of Electrical Engineering, KU Leuven, 3001 Leuven, Belgium\\
\authormark{2}imec vzw, Kapeldreef 75, 3001 Leuven, Belgium\\
\email{\authormark{*}contact@thomasgoossens.be\\ \url{https://orcid.org/0000-0001-7589-5038}}}

\begin{abstract} Vignetting in camera lenses is generally seen as something to avoid.
For spectral cameras with thin-film interference filters, however,  
we argue that vignetting can be an advantage. When illuminated by focused light, the bandwidth of interference filters increases with the chief-ray angle, causing position-dependent smoothing of the spectra. 
We show that vignetting can be used to reduce smoothing and preserve important spectral features.
Furthermore, we demonstrate that by adding additional vignetting to a lens, measurements can be made more consistent across the scene.
This makes vignetting a useful parameter during spectral camera design.
\end{abstract}

\section{Introduction}
\label{sec:orgfca3fd6}
Spectral cameras simultaneously measure spectral and spatial
information. The integration of thin-film interference filters on 
commercially available image sensors enabled very compact and lightweight spectral imaging technology \cite{Tack2012}.

A concern with the integrated thin-film filters is that their transmittance
spectrum depends on the angle of incidence. The larger the incidence
angle, the more the central wavelength shifts
\cite{macleod2017}. Therefore, when a filter is illuminated by an
aperture, the contributions of all the incidence angles in the light
beam need to be added together. This increases the bandwidth (Full
Width at Half Maximum) of the filters and therefore has a position-dependent smoothing effect on the measured spectrum.
This could be problematic for classification or in
applications where localized features of the spectrum are used to
quantify material properties (e.g. vegetation indices \cite{Wu2008}).

For conventional lenses, the larger the object field angle, the larger
the image side chief-ray angle, creating additional shift and
smoothing. To avoid this, telecentric lenses could be used. These lenses, however, strongly limit the available options for camera design and are typically more
expensive. It would therefore be very useful to be able to use
commercially available non-telecentric lenses.    

In our previous work we showed that changes in central wavelength can
be easily corrected in software \cite{Goossens2018,Goossens2019a}. Smoothing of the spectra, however,
cannot be simply reversed in software. In this work we show the smoothing can be controlled by selecting or
designing lenses with optical and mechanical vignetting \cite{Goossens2019a}.

\begin{figure}[h!]
	\centering
\begin{subfigure}[t]{0.49\linewidth}
	\includegraphics[height=0.99\linewidth,page=1]{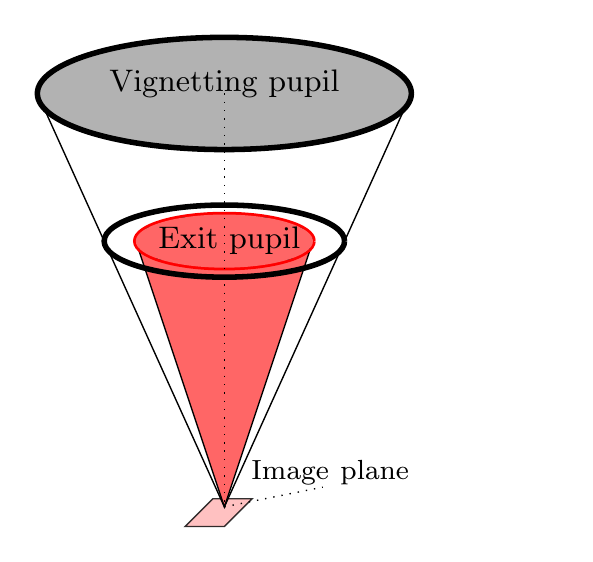}
	\caption{\label{fig:vignetting0} At zero chief ray angle, light is focused from the exit pupil.}
\end{subfigure}
\begin{subfigure}[t]{0.49\linewidth}
		\includegraphics[height=0.99\linewidth,page=2]{./fig/twomodels.pdf}
	\caption{\label{fig:vignetting1} Non-zero chief ray angle. Part of the exit pupil appears cut off.}
\end{subfigure}
\caption{\label{fig:vignetting} Vignetting cuts off part of the light beam for non-zero chief ray angles. The two colored circles are the image-space projections of the most limiting apertures.}
\end{figure}

The key insight is that vignetting cuts off part of the light
beam (Fig. \ref{fig:vignetting}). This can help to control the distribution of incidence angles
and thus control the smoothing.
In this article we thus argue that having a lens with vignetting can
be advantageous. This new insight enables many more lenses to be
seriously considered for use in spectral cameras. We also suggest that
it might be worthwhile to design a lens with deliberate vignetting in
mind. 

We expect our work to be relevant for any system where angle-sensitive
filters are illuminated by a lens. Examples include linear variable
filters \cite{Mu2019a,Renhorn2016,Renhorn2019}, Fabry-Pérot filters, thin-film
filters \cite{Tack2012}, liquid crystal tunable filters
\cite{Stoltzfus2017}. 

In Section \ref{orgac42f73}, we introduce the basic theory to model
vignetting. In Section \ref{org8d28290}, we show how vignetting can be an
advantage in lens selection and how adding additional vignetting can
improve spectral resolution.
\section{Vignetted-aperture theory}
\label{sec:org231102c}
\label{orgac42f73}
Vignetting is the variation of intensity that remains in the image
even when the scene is spatially uniform. The decrease in intensity is caused by the light beam
being physically cut off in the lens. This happens because of the
limited size of a diaphragm, lens or lens hood. There are other forms of vignetting which
are discussed in \cite{Goossens2019a} but will not be considered here.

This cutting off of the light beam changes the distribution of
incidence angles and therefore also the shape of the transmittance spectrum.
In \cite{Goossens2019a}, we showed that the distribution of angles can
be modeled using circle-circle intersections
(Fig. \ref{fig:vignetting1}). Here, part of the exit pupil is cut off
by another circle. While this model works,
no physical interpretation was offered that allows to find the model
parameters from a given lens design. Therefore, an important contribution of this
letter is that, given the lens design, we provide a physically meaningful approach to
model the effect of vignetting.

To make the model physically meaningful, we consider the projection of
each lens component into the image space of the lens. Evidently, each such image will have a certain radius
and position. The \emph{exit pupil} is the image of the component that most
limits the solid angle of the focused light at zero chief ray angle. The second
most limiting aperture we call the \emph{vignetting pupil} (Fig. \ref{fig:vignetting}).
By construction, each pupil represents a limitation on the rays that
can pass through a lens; Thus, only light rays
seemingly coming from both the exit pupil and the vignetting pupil can
occur. Thus, for increasing chief-ray angles, vignetting happens because the vignetting pupil starts to limit
the available light rays (Fig. \ref{fig:vignetting1}).

This physical interpretation in image space explains the use of the circle-circle
intersection model. The circle that cuts off part
of exit pupil is obtained by projecting the vignetting pupil to
the same plane as the exit pupil (Fig. \ref{fig:vignetting1}). This is
further explained in Appendix \ref{orgae4c2d0}.

\section{Vignetting as an advantage}
\label{sec:org393b2ee}
\label{org8d28290}
\subsection{Removing vignetting}
\label{sec:orgfba1d87}

In this section we demonstrate that removing the vignetting of an
existing lens can have an important position-dependent smoothing effect.
We make use of the Edmund Optics C Series 16 mm lens at \(f/1.6\). This lens has
significant vignetting at this f-number and this was described and modeled in
\cite{Goossens2019a}. Here, we will use the same methods to simulate
the effect of vignetting.

We simulated the measurement of the transmittance of a Thorlabs FGB67S Colored
Glass Bandpass filter. The
transmittance \(\boldsymbol{\tau}\) is obtained using a flat-fielding approach
such that each element is calculated as
\begin{equation}
\tau_i = \dfrac{\text{Output with filter}}{\text{Output without filter}} = \dfrac{\int_\Lambda F(\lambda) T_i(\lambda) S(\lambda) \dd\lambda}{\int_\Lambda T_i(\lambda) S(\lambda) \dd\lambda}.
\end{equation}
Here \(S(\lambda)\) is the source spectrum, \(F(\lambda)\) the
transmittance of the colored glass and \(T_i(\lambda)\) is the
transmittance of a filter on the pixel array with index \(i\). The
calculation of the correct filter response is briefly discussed in
Appendix \ref{org6982521}. 
The measured transmittance \(\boldsymbol{\tau}\) is then compared for different chief ray
angles (i.e. positions in the scene) and is corrected for the shift in central
wavelength (See Appendix \ref{org6982521}).

\begin{figure}[H]
\begin{subfigure}[t]{0.49\linewidth}
\includegraphics[width=0.99\linewidth]{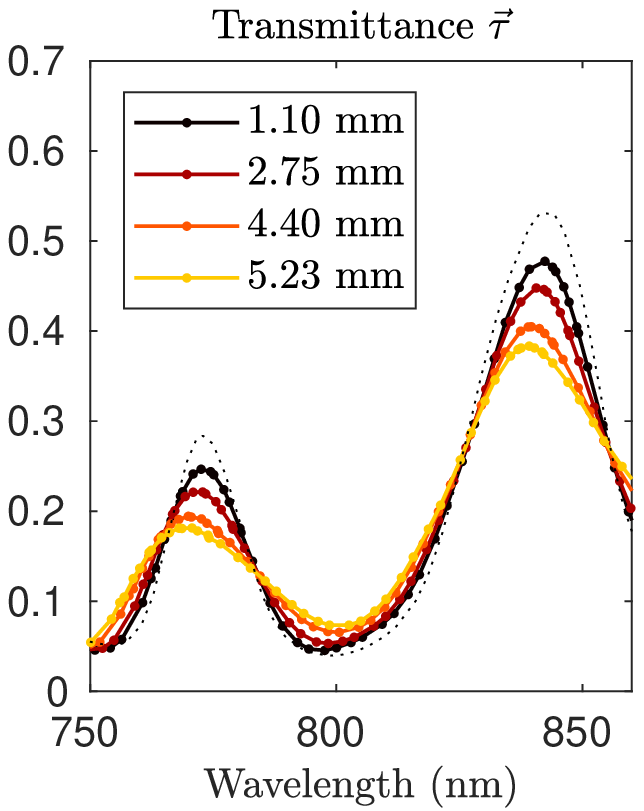}
\caption{\label{fig:eo16} Vignetting removed}
\end{subfigure}
\begin{subfigure}[t]{0.49\linewidth}
\includegraphics[width=0.99\linewidth]{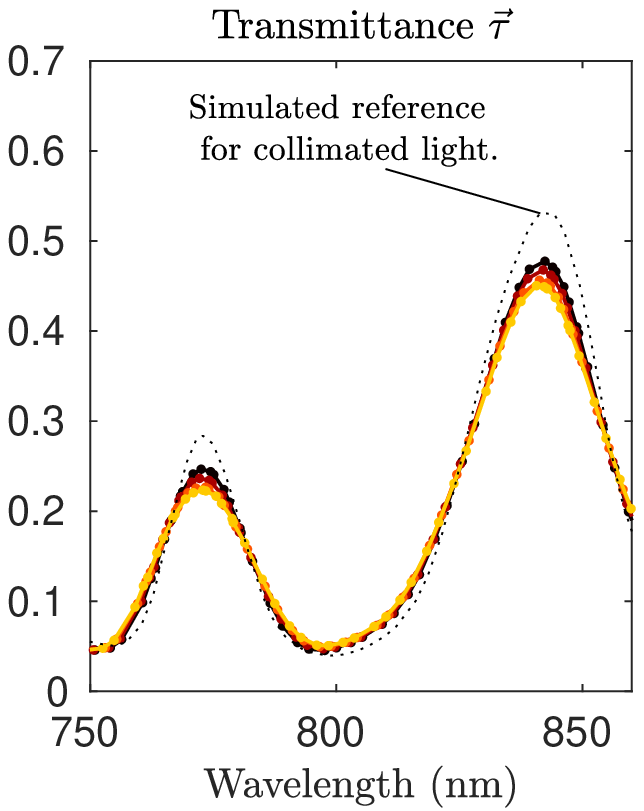}
\caption{\label{fig:eo16vignet} With vignetting.}
\end{subfigure}
\caption{\label{fig:eo16sim} \textbf{Simulated response with FGB67 filter at $\mathbf{f/1.6}$.} If this 16 mm lens had no vignetting, the spectrum would be severely smoothed. Vignetting thus can improve the consistency of the measurements at different positions. The distances are measured from the optical axis in the image plane. The spectra are corrected for shifts (see Appendix C).}
\end{figure}

When we simulate this, but with vignetting removed, the
resolution of the measured spectrum severely deteriorates (Fig. \ref{fig:eo16}). 
A lens should therefore not be discarded just because it has some vignetting.
Vignetting should rather be seen as a design parameter that
can be taken into account when selecting or designing a lens for a
thin-film spectral camera.

Fig. \ref{fig:eo16} also shows the simulated response for collimated
since this effect will be more severe for any converging beam. It is
used as a reference since this is the minimal smoothing than can be
obtained using vignetting.  

\subsection{Engineered vignetting}
\label{sec:orgd0c390f}
In this section we demonstrate that vignetting can be added to an
existing lens to control the spectral smoothing at different positions. This experiment is meant as a
proof-of-concept that vignetting can be kept in mind when designing lenses for spectral imaging.

Without loss of generality, we found no lens with the same f-number, focal length, no vignetting \emph{and} a
comparable exit pupil distance as the Edmund Optics lens 
from the previous section. Nevertheless, we found a high quality lens
(Optec OB-V-SWIR 16mm C1326) that has no vignetting, thus making it suitable for this experiment. The exit pupil
distance is 39.37 mm (instead of 21 mm). This implies smaller chief
ray angles, and therefore already less smoothing.

\begin{figure}[H]
\centering
	\includegraphics[width=0.7\linewidth]{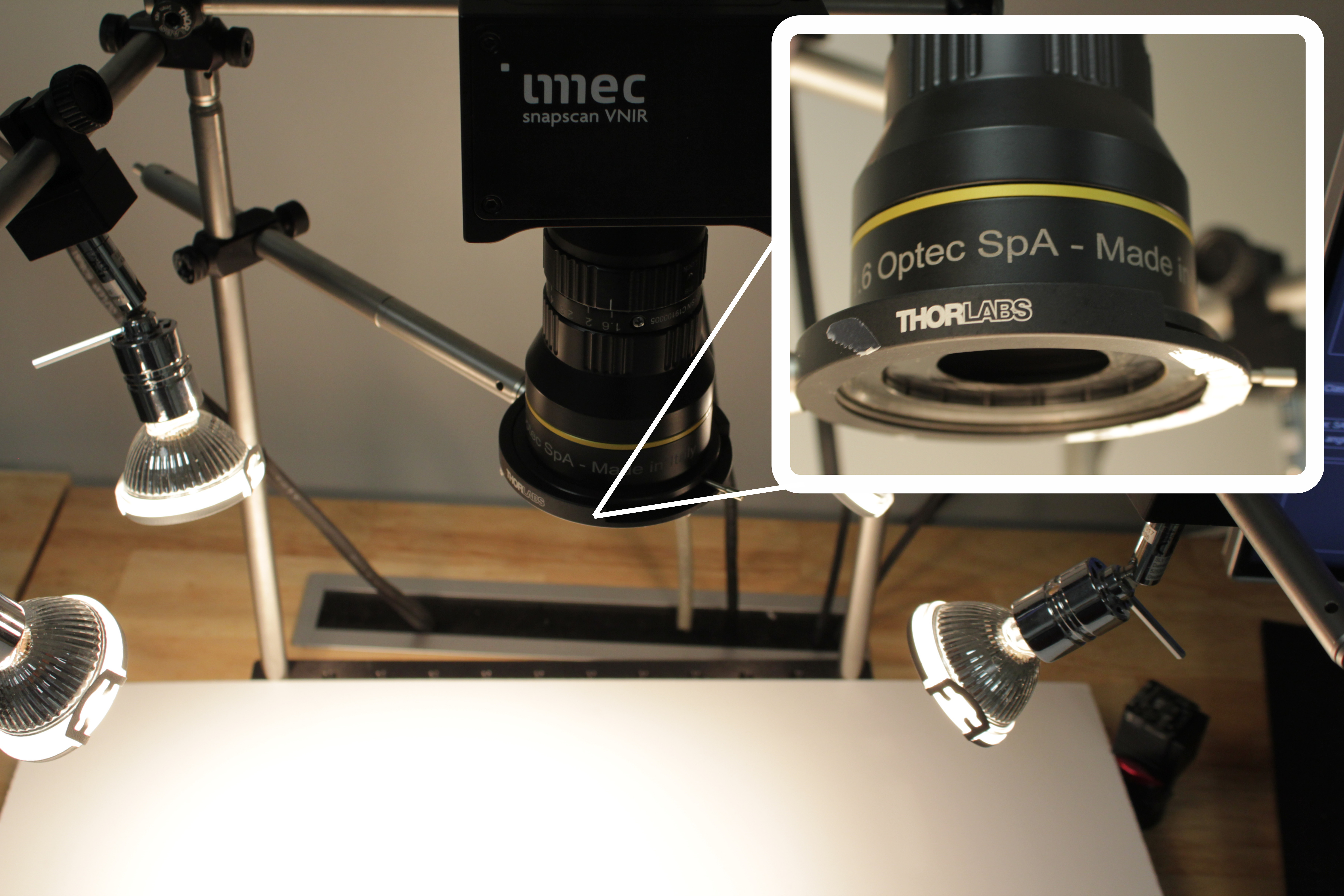}
	\caption{\label{fig:aperture} Introducing vignetting by placing an aperture in front of a lens.}
\end{figure}

In this experiment we measure the transmittance spectrum of the Thorlabs FGB67S Colored
Glass Bandpass filter using imec's Snapscan VNIR camera
\cite{Pichette2017a}. Analogous to the simulation, the measured spectra are compared
for several positions in the scene. The experiment is then repeated
with a 30 mm wide circular aperture placed in front of the lens
(Fig. \ref{fig:aperture}). 

\begin{figure}[H]
\begin{subfigure}[t]{0.49\linewidth}
\includegraphics[width=0.99\linewidth]{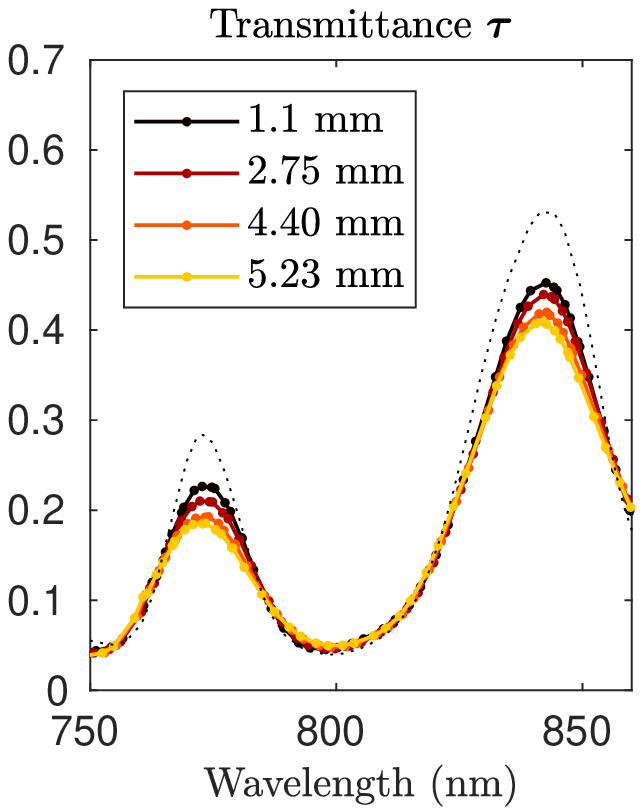}
\caption{\label{fig:optec16real} No aperture}
\end{subfigure}
\begin{subfigure}[t]{0.49\linewidth}
\includegraphics[width=0.99\linewidth]{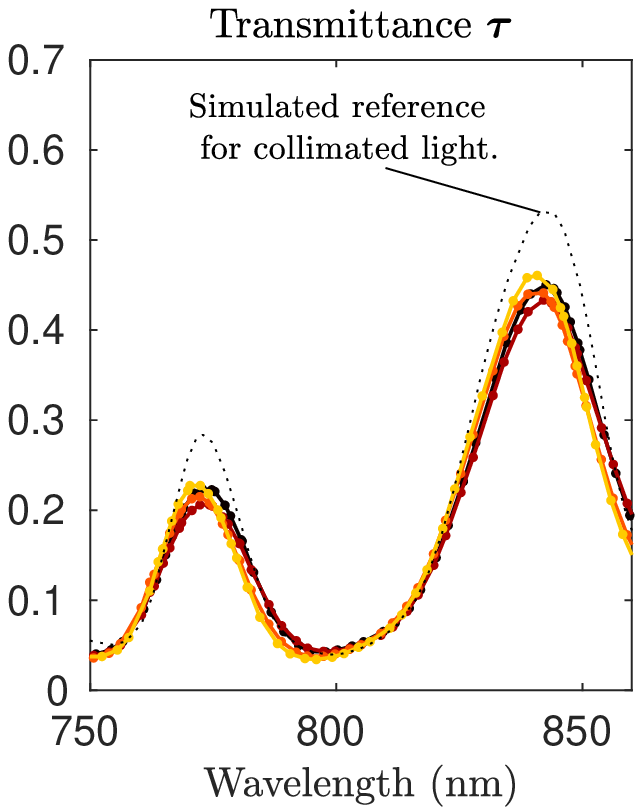}
\caption{\label{fig:optec16vignetreal} With aperture.}
\end{subfigure}
\caption{\label{fig:advantagereal} \textbf{Real measurement.} Because the used lens has almost no vignetting, the increased filter bandwidth causes more smoothing of the spectra (Fig. \ref{fig:fwhmmap}). This smoothing can be controlled by placing an aperture in front of the lens but not reduced more than the reference.}
\end{figure}

By adding the chosen aperture, the shape of the spectrum
becomes more consistent for different positions in the scene (Fig. \ref{fig:advantagereal}).
The effect is smaller than the simulation of Fig. \ref{fig:eo16sim}
due to the longer exit pupil distance which implies smaller chief-ray angles.

\begin{figure}[H]
\centering
	\includegraphics[width=0.9\linewidth]{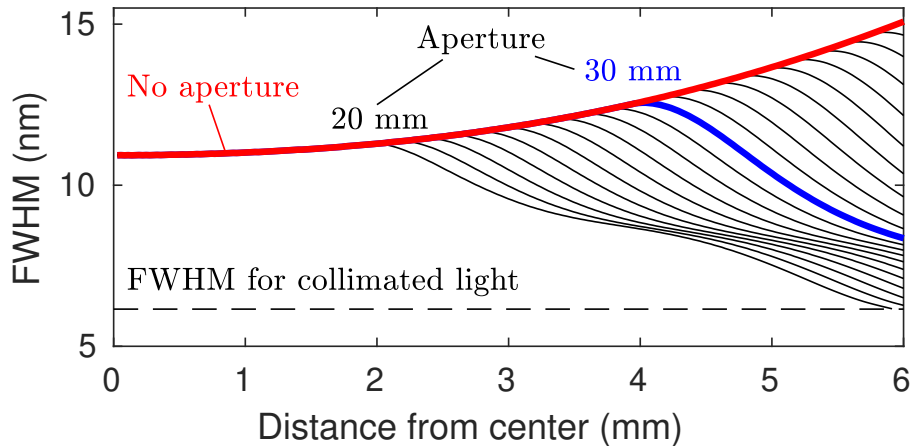}
\caption{\label{fig:fwhmmap} Using simulation, the variation of the FWHM can be explored for different aperture diameters (shown in steps of 1 mm) for $f/1.6$. The 30 mm aperture was used for the real experiment.}
\end{figure}

To compare the effect of different apertures, the variation of the
Full Width at Half Maximum (FWHM) of the filters can be simulated
(Fig. \ref{fig:fwhmmap}). This is done by calculating the filter
transmittance \(T(\lambda)\) and its FWHM for each position in the scene
(Appendix \ref{org6982521}).  
The optimal choice of the aperture dimensions can then be made
according to some chosen criterion.
Sometimes one might want to minimize the FWHM as much as possible.
In other cases, one might prefer a more constant FWHM for the
given field of view. 
For example, because the FWHM at \(d=5.23\) mm is slightly smaller than the FWHM at 1.1 mm
(Fig. \ref{fig:fwhmmap}), the smoothing at 5.23 mm is reversed slightly
more. This explain why the this measurement has a slightly higher peak
value in Fig. \ref{fig:optec16vignetreal}. The smallest FWHM that can be achieved is the FWHM of the filter
measured with collimated light.

\subsection{Trade-off with signal-to-noise ratio}
\label{sec:org3ceef4d}
We have shown that vignetting can be an advantage for spectral
imaging. The trade-off is that light is being cut-off which decreases
the signal-to-noise (SNR) ratio and increases the spatial non-uniformity
of the light in the image. 

\begin{figure}[H]
\includegraphics[width=0.99\linewidth]{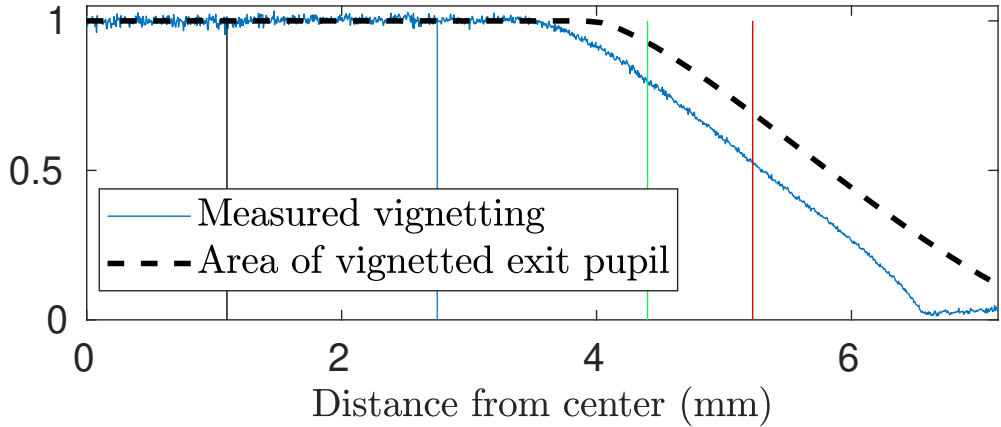}
\caption{\label{fig:profile} Measured vignetting profile versus predicted change. Both quantities are normalized by their value at the central position (zero chief-ray angle). The vertical lines indicate the distances at which the spectra were sampled (Fig. \ref{fig:advantagereal}).}
\end{figure}

The decrease in SNR can be modeled by calculating the area
of the vignetted exit pupil using circle-circle intersections
(Fig. \ref{fig:vignetting1}). The area changes discontinuously when the vignetting pupil abruptly starts cutting.
The predicted vignetting slightly misses the onset point of vignetting
(Fig. \ref{fig:profile}). This difference can be explained by a 1 to 2
mm measurement error on the aperture dimensions, which is
realistic for the given setup. More accurate calibration can be obtained using the method
from \cite{Goossens2019a}.

The exact trade-off will depend on the application requirements, the
system and the available illumination in the setup.

\section{Conclusion}
\label{sec:orga4c4b55}

We have shown that vignetting is an important design parameter for
spectral cameras with (thin-film) interference filters. 
When selecting the camera/lens combination to match application
requirements, one should consider how vignetting can help influence
the smoothing of spectrum. A trade-off must then be made with the loss in
signal-to-noise ratio. The optimal design parameters can be explored
using simulation.

It is also possible to engineer vignetting by either adding an
aperture to an existing lens or design a lens from scratch.
The vignetting may be engineered to minimize smoothing but also
to make the measured spectra of identical samples more \emph{consistent} across the scene. 

\section{Acknowledgements}
\label{sec:org2480aa4}
We would like to thank Massimiliano Musazzi and Optec spa for their
kind assistance in providing the lens and the necessary design
parameters. We also thank Prof. Claude Amra and Ruben
Van De Vijver for the helpful discussions.
\section{Disclosures}
\label{sec:org0514926}
The authors declare no conflict of interest.

\appendix
\section*{Appendix}

In these appendices we discuss how to calculate the effect of adding an
additional aperture in front of an existing lens. The steps will be to
calculate the position of its image (the vignetting pupil) and use
that to calculate the model parameters of the circle-circle
intersection model so that the theory from \cite{Goossens2019a} can be
applied for simulation and shift correction.

\section{Calculating the dimensions of the vignetting pupil}
\label{sec:org8374424}
\label{org92acdf2}

Let us place an aperture of radius \(Q_\text{obj}\) on the object side of the
lens. Its position \(z_\text{obj}\) is measured from the front focal plane. If
the front focal plane is positioned within the lens, \(z_\text{obj}\)
will always be negative by construction. The positions of focal planes
can usually be found in a datasheet or lens design file \cite{OptecS.p.a2014}. 

We calculate the position \(z_\text{im}\) and radius \(Q_\text{im}\) of the image of this aperture using the Newtonian equations
\cite{greivenkamp2014}. As \(z_\text{im}\) is measured from
the rear focal plane it follows that
\begin{equation}
z_\text{im} = -\dfrac{f_\text{eff}^2}{z_\text{obj}},
\end{equation}
where \(f_\text{eff}\) is the effective focal length.

The radius of the vignetting pupil is then found to be
\begin{equation}
Q_\text{im}= -\dfrac{f_\text{eff}}{z_\text{obj}}Q_\text{obj}.
\end{equation}

\begin{figure}[htbp!]
\centering
	\includegraphics[width=0.9\linewidth]{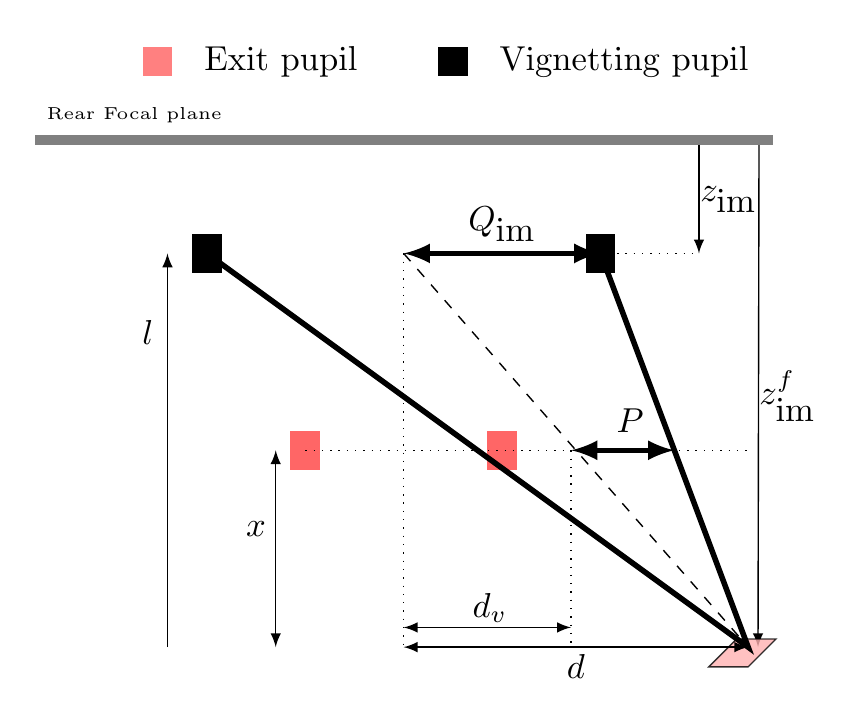}
\caption{\label{fig:project} \textbf{A slice of Fig. \ref{fig:vignetting}}. Projecting the vignetting pupil of radius $Q_\text{im}$ onto the exit pupil plane gives a circle of radius $P$.}
\end{figure}

Let us define \(z_\text{im}^f\) as the the position of the rear focal plane as measured from the
image plane. The position \(l\) of the vignetting pupil, as measured from the image
plane then becomes \(l = -(z_\text{im}-z_\text{im}^f)\) (Fig. \ref{fig:project}).

In the experiment shown in Fig. \ref{fig:aperture}, the following
values were calculated using the lens design files provided by the
manufacturer and the dimensions of the aperture: \(f_\text{eff} = 16.1\,
\text{mm}\), \(z_\text{im}^f=0.072\,\text{mm}\), \(z_\text{obj}=-34.207\, \text{mm}\), \(Q_\text{obj}=\frac{30}{2}\,\text{mm}\) and thus \(z_\text{im}=7.578\, \text{mm}\).

\section{Transforming to the circle-intersection model}
\label{sec:orga854857}
\label{orgae4c2d0}

The circle-intersection model from \cite{Goossens2019a} describes how
vignetting occurs by the exit pupil being partially cut off by another
circle (Fig. \ref{fig:vignetting1}). Two model parameters are required: The radius \(P\) of the circle that cuts off part of the
exit pupil, and a sensitivity parameter \(h\) which is defined such that
the position \(d_v\) of the circle behaves as 
\begin{equation}
d_v = h\tan\cra.
\end{equation}
The chief ray angle \(\cra\) is defined as \(\tan \cra = \frac{d}{x}\)
with \(x\) being the distance to the exit pupil
(Fig. \ref{fig:project}).

As discussed earlier, the circle-intersection model is related to the
vignetting pupil by a projection (Fig. \ref{fig:vignetting1} and \ref{fig:project}).
Therefore, using elementary geometry, the radius \(P\) of the projected circle can be found to be 
\begin{equation}
P = \left|\dfrac{x}{l} \right| Q_\text{im}
\end{equation}
and its position
\begin{equation}
d_v = \left(1-\dfrac{x}{l} \right) x \tan\cra = \underbrace{\left(x-\dfrac{x^2}{l} \right)}_{h} \tan\cra,
\end{equation}
such that \(h = \left(x-\frac{x^2}{l}\right)\).
The absolute value takes into account that \(l\) can, in principle, be negative. 

For the experiment in Fig. \ref{fig:aperture}, we calculated that \(P =
37.03\) mm and \(h = 245.9\) mm.

\section{Simulating the circle-intersection model}
\label{sec:org6b151ec}
\label{org6982521}
This section briefly summarizes the key aspects on how the circle-intersection model can
be applied. For a detailed analysis, we refer to \cite{Goossens2019a}.

\begin{figure}[H]
\begin{subfigure}[t]{0.49\linewidth}
\includegraphics[width=0.99\linewidth]{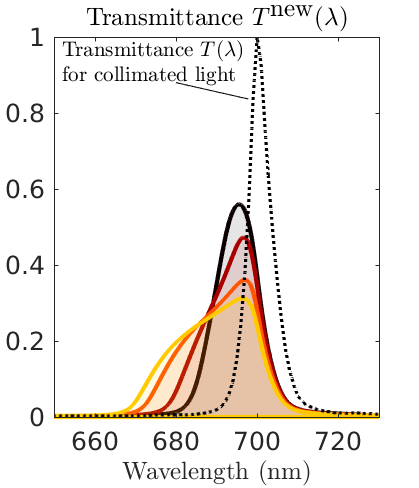}
\caption{\label{fig:filterT}Filter transmittance $T^\text{new}(\lambda)$ calculated using kernel convolution.}
\end{subfigure}
\begin{subfigure}[t]{0.49\linewidth}
\includegraphics[width=0.99\linewidth]{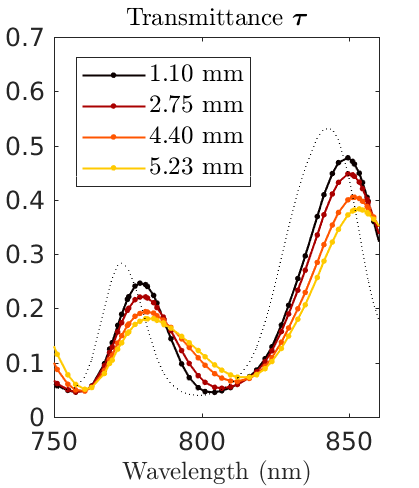}
\caption{\label{fig:shifted}Measured spectra.}
\end{subfigure}
\caption{\label{fig:nocwl}\textbf{Simulated measurement with FGB67 filter at $\mathbf{f/1.6}$.} For increasing chief-ray angle, the position and FWHM of the filter transmittance changes (\subref{fig:filterT}). This causes the measured spectra to be shifted as well (\subref{fig:shifted}). This can be corrected using \eqrefn{eq:vignetcorrect} as in Fig. \ref{fig:eo16}.}
\end{figure}

Let \(T(\lambda)\) be the transmittance of a filter illuminated with collimated
light at normal incidence. 
When such a filter is irradiated by focused light, its effective
transmittance \(T^{\text{new}}(\lambda)\) will be a shifted and smoothed
version of \(T(\lambda)\) (Fig. \ref{fig:filterT}).

In \cite{Goossens2018,Goossens2019a}, this is modeled by a convolution operation on \(T(\lambda)\) such that
\begin{equation}
T^{\text{new}}(\lambda) = (k*T)(\lambda) = \int_{-\infty}^\infty k(\delta_\lambda) T(\lambda-\delta_\lambda) \textup{d}\delta_\lambda.
\end{equation}
The kernel \(k(\delta_\lambda)\) encodes the change in central
wavelength and bandwidth for a given distribution of incidence angles. A closed-form formula for
\(k(\delta_\lambda)\) can be found in \cite{Goossens2019a}, but will not
be further discussed here.
It suffices to know that for a lens with vignetting, this kernel is completely defined by four parameters:
the radius \(R\) and position \(x\) of the exit pupil, and the radius \(P\)
and horizontal position \(d_v\) of the vignetting circle that intersects the exit pupil (Fig. \ref{fig:project}).
 
The spectra are displayed by plotting each response $\tau_i$ at its correspending central wavelength. These central wavelengths were determined for collimated light at normal incidence. 
Because the actual central wavelengths are shifted, the measured spectra will also
be shifted (Fig. \ref{fig:shifted}). 
These shifts can be corrected by updating the wavelength at which each
filter response is plotted such that
\begin{equation}
\label{eq:vignetcorrect}
\lcwl^\text{new} = \lcwl + \int_\Lambda \lambda k(\delta_\lambda)\textup{d}\delta_\lambda,
\end{equation}
where the shift is calculated as the mean value of the kernel. This
approach is used in \cite{Goossens2018,Goossens2019a} and further
justified in the discussion of \cite{Goossens2020}.

\bibliography{/home/thomas/Documents/imec/phd/library/library.bib}
\end{document}